\documentclass[twocolumn,groupedaddress,preprintnumbers,amsmath,amsfonts,amssymb,aps,floatfix,nofootinbib,superscriptaddress]{revtex4}
\usepackage{graphicx}
\usepackage{hyperref}

\newcommand{\ket}[1]{|{#1}\rangle}
\newcommand{\bra}[1]{\langle{#1}|}

\newcommand{\beq}{\begin{equation}}
\newcommand{\eeq}{\end{equation}}
\newcommand{\bea}{\begin{eqnarray}}
\newcommand{\eea}{\end{eqnarray}}
\newcommand{\grad}{\vec\nabla}
\newcommand{\xprime}{\vec x\hspace{1pt}'}

\begin{document}

\title{Trapping ions with lasers}

% \author{Cecilia Cormick$^1$, Tobias Schaetz,$^2$ and Giovanna Morigi$^1$}
% \affiliation{$^1$Theoretische Physik, Universit\"at des Saarlandes, D-66041 Saarbr\"ucken, Germany\\
% $2$ Max-Planck-Institut f\"ur Quantenoptik, Hans-Kopfermann-Stra{\ss}e 1, D-85748 Garching, Germany}
% \email{cecilia.cormick@physik.uni-saarland.de}

\author{Cecilia Cormick}
\email{cecilia.cormick@physik.uni-saarland.de}
\affiliation{Theoretische Physik, Universit\"at des Saarlandes, D-66041 Saarbr\"ucken, Germany}

\author{Tobias Schaetz}
\affiliation{Max-Planck-Institut f\"ur Quantenoptik, Hans-Kopfermann-Stra{\ss}e 1, D-85748 Garching, Germany}

\author{Giovanna Morigi}
\affiliation{Theoretische Physik, Universit\"at des Saarlandes, D-66041 Saarbr\"ucken, Germany}

\date{\today}

\begin{abstract}
This work theoretically addresses the trapping of an ionized atom with a single valence electron by means of lasers, analyzing qualitatively and quantitatively the consequences of the net charge of the particle. In our model, the coupling between the ion and the electromagnetic field includes the charge monopole and the internal dipole, within a multipolar expansion of the interaction Hamiltonian. Specifically, we perform a Power-Zienau-Woolley transformation, taking into account the motion of the center of mass. The net charge produces a correction in the atomic dipole which is of order $m_e/M$ with $m_e$ the electron mass and $M$ the total mass of the ion. With respect to neutral atoms, there is also an extra coupling to the laser field which can be approximated by that of the monopole located at the position of the center of mass. These additional effects, however, are shown to be very small compared to the dominant dipolar trapping term.
\end{abstract}

\maketitle

\section{Introduction}

Charged particles are routinely trapped in several laboratories worldwide~\cite{BOOK_Ghosh_1995,Dubin_O'N_1999}. Confinement is usually achieved by means of radiofrequency (rf) traps or by combining static electric and magnetic fields \cite{BOOK_Ghosh_1995}. These technologies have reached such a level of precision that they are now at the basis of mass spectrometers~\cite{Kienle_2001}, optical clocks~\cite{Roos_C_K_R_B_2006, Rosenband_et_al_2008}, and prototypes of quantum information processors \cite{Haeffner_R_B_2008,Home_et_al_2009, Monz_et_al_2009}.

The creation of optical potentials for single ions by means of laser beams \cite{Katori_S_W_1997, Schlipf_K_P_W_1998} has been discussed with the purpose of employing established techniques routinely used to trap neutral atoms~\cite{Grimm_W_O_1999}. Recently, in Ref.~\cite{Schneider_E_H_S_2010} the trapping of a single ion in a dipole trap was reported. This alternative approach to confine ions optically is expected, among other aims, to allow for the study and realization of controlled dynamics coupling atoms and ions~\cite{Calarco2007}, thereby overcoming limitations which could arise from the field of ion traps~\cite{Grier_C_O_V_2009, Zipkes_P_S_K_2010, Schmid_2010, Zipkes_R_S_K_2010}. However, in comparison to neutral atoms, ions have a net charge which additionally couples to the electromagnetic field. Thus, one should evaluate the impact of the oscillation of the ion in the optical potential, and the direct coupling of the charge to the time-varying electromagnetic field. In common rf traps, the latter leads to a direct drive of the ion's motion at the rf frequency -- an effect called micromotion \cite{BOOK_Ghosh_1995} that can severely affect experiments that require not only low temperatures but also small residual kinetic energies. Even though the physical principles of radiofrequency traps and optical traps are similar, the frequency of the electromagnetic field differs by eight orders of magnitude and therefore leads to dramatically different weights for the amplitudes of, in principle, similar effects. 

\begin{figure}[ht]
\begin{center}
\includegraphics[width=0.35\textwidth]{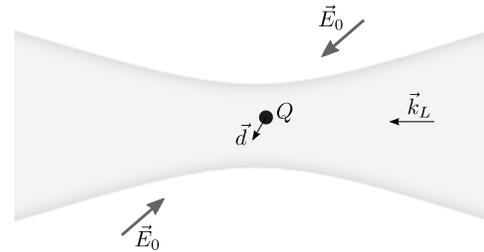}
\caption{\label{fig:trap} An ion is confined in a dipolar trap formed by a focused red-detuned laser beam, with wave vector $\vec k_L$. The particle couples with the external fields both via the charge monopole $Q$ and the optical dipole $\vec d$. The first coupling gives rise to an effect which is similar to micromotion in an rf trap; the latter is responsible for the optical trapping. An electrostatic field $\vec E_0$ can be used to prevent the escape of the particle along the propagation direction of the laser, as realized in~\cite{Schneider_E_H_S_2010}.} 
\end{center}
\end{figure}

Starting from this motivation, in this work we analyze theoretically the efficiency of laser trapping of an alkali-earth metal ion, namely, an ionized atom with a single valence electron. We start from the Hamiltonian in Coulomb gauge, treating the electromagnetic field in second quantization and the atom in the regime in which relativistic effects are perturbative corrections. We then perform the Power-Zienau-Woolley transformation, taking into account the quantized center-of-mass motion. This allows us to identify a multipole expansion, in which the dominant term is responsible for dipolar trapping. We analyze the effect and estimate the order of magnitude of the corrections, specially focusing on the terms which originate from the charge monopole. In the spirit of \cite{Glauber_1992}, reporting a study on the stability of a single ion in a Paul trap, in our work we perform a stability analysis of an ion in a dipole trap, where an effect similar to micromotion emerges from the coupling of the ion charge with the fast-oscillating laser field, and we briefly discuss the impact on potential applications. For definiteness, we shall consider the setup drawn in Fig.~\ref{fig:trap}, but our results remain valid for ions trapped in optical lattices. 

This paper is organized as follows: In Sec. \ref{Sec:dipolar trapping} the basic physical concepts underlying the dipolar trapping of particles are briefly reviewed. In Sec.~\ref{Sec:ion-photon interactions} the Hamiltonian in Coulomb gauge is introduced and the Power-Zienau-Woolley transformation is performed. The various multipolar coupling terms emerging from the transformation are discussed in Sec.~\ref{Sec:corrections}, while concluding remarks are presented in Sec.~\ref{Sec:conclusions}.

\section{Dipolar trapping}
\label{Sec:dipolar trapping}

We first briefly summarize the concepts at the basis of dipolar trapping of atoms (we refer the reader to \cite{Grimm_W_O_1999} for a review). We assume an alkali-metal atom. For sufficiently weak laser fields, the coupling between the atom and the laser field only involves the valence electron, and can be described by the interaction Hamiltonian in the electric dipole approximation, which reads
\beq
H_{\rm dip} = -\vec d \cdot \vec E (\vec R) \label{eq:dipolar}
\eeq
where $\vec d$ is the atomic dipole and $\vec E (\vec R)$ is the electric field corresponding to the trapping laser, evaluated at the position $\vec R$ of the atomic center of mass. In the following we will assume that the atomic motion has been brought to energies corresponding to temperatures of the order of the millikelvin by means of laser cooling~\cite{Stenholm_1986,Wineland1978}. In this regime, the characteristic time scale for the evolution of the atomic center of mass is usually much longer than the time scale in which the electronic degrees of freedom evolve towards the internal steady state. One can then evaluate the mechanical effects on the atom by means of an adiabatic treatment, namely, assuming that the internal state is the stationary state corresponding to the field at position $\vec R$, and which results from the competition of the dipolar excitation due to the laser and the relaxation due to spontaneous decay~\cite{Stenholm_1986}. 

\begin{figure}[ht]
\begin{center}
\includegraphics[width=0.4\textwidth]{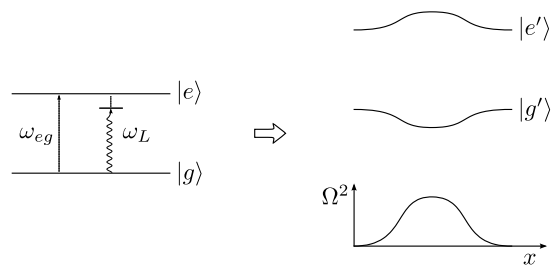}
\caption{\label{fig:levels} Left panel: Optical trapping is achieved by means of a laser that drives a dipolar transition between a ground state $\ket{g}$ and an excited state $\ket{e}$. The radiation is detuned by an amount $\delta = \omega_L - \omega_{eg}$ with respect to the atomic transition. Right panel: In the dressed atom picture, corresponding to the eigenstates of the atomic Hamiltonian including the coupling to the classical laser field, the levels have energy shifts that depend on the Rabi frequency $\Omega$. This gives rise to an effective potential associated with the spatial variation of the laser intensity, which is proportional to $\Omega^2$ and is here reported along a direction $x$ transverse to the propagation direction of the laser.}
\end{center}
\end{figure}

We consider thus a laser which drives a transition between a ground state $\ket{g}$ and an excited state $\ket{e}$, forming an optical dipole transition at frequency $\omega_{eg}$. The energy diagram is sketched in Fig.~\ref{fig:levels}. The laser is a classical field, more specifically, a travelling wave with frequency $\omega_L$, wave  vector $\vec{k}_L$, polarization $\hat{e}_L$, and amplitude $E_L({\vec R})$, corresponding to a tightly focussed beam. From the solution for the internal steady state of the atom as a function of $\vec R$, one can find the mechanical force governing the atomic center-of-mass motion and which reads~\cite{Nienhuis_vdS_S_1991,API_Cohen-Tannoudji_D-R_G_1998}:
\beq
\langle \vec F\rangle = - \frac{\hbar\delta}{2} \vec\nabla \ln(1+s) + \frac{\hbar\Gamma}{2} \frac{s}{s+1} \vec k_L \label{eq:force}
\eeq
(after averaging over the optical oscillation period). Here, $s=s(\vec{R})$ is the saturation parameter, which depends on the position through the spatial dependence of the Rabi frequency, and reads
\beq \label{eq:s}
s(\vec{R}) = \frac{\Omega^2(\vec{R})/2}{\delta^2+\Gamma^2/4},
\eeq
with $\Gamma$ the natural linewidth of the dipolar transition, $\delta = \omega_L - \omega_{eg}$ the detuning of the laser frequency from resonance, and 
\beq\label{eq:Rabi}
\Omega(\vec{R}) = \frac{1}{\hbar} \left|\bra{g}\vec d\ket{e} \cdot \hat e_L E_L(\vec{R})\right|
\eeq
the Rabi frequency.

The first term in Eq.~(\ref{eq:force}) describes a dispersive force which emerges from the intensity gradient, namely, the so-called dipolar force. The second term is the radiation-pressure force, which is related to dissipative processes. The strength of the dipolar force can exceed by orders of magnitude the radiation pressure when the atom is weakly driven, namely, when the saturation parameter satisfies $s\ll 1$, and when the laser is far detuned from the transition, such that $|\delta|\gg\Gamma/2$ (provided that the field is still far-off resonance from other electronic states). In this limit, the mechanical effects due to the radiation pressure can be neglected and the effect of the dipolar coupling can be described by the dipole force. This leads to the definition of an effective conservative potential for the center of mass, which is proportional to the laser intensity and takes the form
\beq
V_{\rm eff}(\vec{R}) \simeq \frac{\hbar\delta}{2} s(\vec R)\,. \label{eq:effective potential}
\eeq
(where the approximation $s\ll1$ has been used). Since $s(\vec R)$ is positive, the sign of the potential is solely determined by the detuning $\delta$. When $\delta <0$, the maxima of the field intensity are minima of the effective potential, and the atom can be trapped in high-intensity regions. In a dressed-atom picture, the effective potential can be identified with the position-dependent energy shift of the atomic levels, as displayed in Fig.~\ref{fig:levels}.

Radiation pressure can slightly shift the equilibrium position with respect to the minimum of the effective potential. Moreover, scattering of photons by the atom leads to heating of the atomic motion by an amount of the order of the recoil energy $E_{\rm rec} = (\hbar k_L)^2/2M$ per scattering event. The rate at which photons are scattered is given by \cite{Grimm_W_O_1999} 
\beq
\Gamma_{\rm sc} \simeq \frac{\Gamma}{\delta} \frac{V_{\rm eff}}{\hbar}, \label{eq:scattering}
\eeq
where the formula is valid at low saturation. For $|\delta|\gg\Gamma/2$, a time scale of the dynamics can then be identified, in which atomic motion is essentially determined by the conservative potential $V_{\rm eff}$, while incoherent scattering of laser photons can be neglected. Further orders of magnitude in the ratio between potential depth and heating may be gained by using a blue-detuned standing wave laser, such that the atoms are confined in the nodes of the field, as in dispersive optical lattices~\cite{BlochRMP}.

In the following we analyze how these dynamics are modified when the atom has a net charge which also couples to the rapidly oscillating electromagnetic field.

\section{Ion-Photon interactions}
\label{Sec:ion-photon interactions}

The system we shall consider is an ionized alkali-earth metal atom, with mass $M$ and total charge $Q$. We treat the ion as a hydrogen-like system, since the radiation which illuminates it is optical and at sufficiently low intensity, such that the induced electronic transitions involve only the valence electron. Center-of-mass and electronic excitations will be treated systematically in the non-relativistic limit. We will focus on the mechanical effects of light on the center-of-mass motion, considering both the coupling of the field with the charge monopole and with the atomic dipole.

For the sake of simplicity we shall take a spinless nucleus as in the experiment reported in \cite{Schneider_E_H_S_2010}, therefore not considering any hyperfine structure~\footnote{In general, due to the large mass difference between electron and nucleus, the coupling of the magnetic field to the nuclear spin is much weaker than that between the magnetic field and the spin of the electron, so that it gives rise to corrections of higher order than the ones reported here.}.

\subsection{Hamiltonian in Coulomb gauge}

We assume a hydrogen-like ion with a core of mass $m_n$ and charge $q_n$, and we denote by $m_e$ and $q_e$ the mass and charge of the valence electron respectively. The total charge is thus $Q = q_n + q_e$, and the total mass is $M = m_n + m_e$. In the non-relativistic regime the Hamiltonian for the atom and the quantum electromagnetic field in the Coulomb (transverse) gauge takes the form
\begin{multline}
H = \sum_{j=n,e} \frac{1}{2 m_j} \left[\vec p_j - q_j\vec A(\vec r_j)\right]^2 + \sum_{j=n,e} q_j \Phi(\vec r_j) + \\
+ V_{\rm Coul}+H_{\rm rel}+ H_{\rm rad} \label{eq:first Hamiltonian}
\end{multline}
where $\vec r_j$ and $\vec p_j$ ($j=n,e$) are conjugate position and momentum, $\vec A$ is the vector potential in the Coulomb gauge,  $V_{\rm Coul}$ contains the Coulomb interaction between core and electron, $\Phi$ represents an external electrostatic potential, $H_{\rm rel}$ gives the relativistic corrections to the atomic Hamiltonian, and $H_{\rm rad}$ is the energy of the electromagnetic radiation in vacuum,
\beq
\label{H:rad}
H_{\rm rad}=\sum_m\hbar\omega_ma_m^\dagger a_m. 
\eeq
Here the subscript $m$ runs over the modes of the electromagnetic field, characterized by the frequency $\omega_m$, wave vector $\vec k_m$, and polarization given by the unit vector $\hat e_m\perp \vec k_m$. The operators $a_m$ and $a_m^{\dagger}$ annihilate and create, respectively, a photon in mode $m$ and satisfy the commutation relations $[a_m,a_{m'}^{\dagger}]=\delta_{m,m'}$. In Eq.~(\ref{H:rad}) we have taken a quantization volume in a cubic box of side $L$ with periodic boundary conditions, and we have set to zero the energy of the vacuum state. In terms of the creation and annihilation operators, the vector potential is given by:
\beq
\vec A (\vec x) = \sum_m \sqrt{\frac{\hbar}{(2\epsilon_0\omega_mL^3)}} \left[ a_m \hat e_m e^{i\vec k_m \cdot \vec x} + {\rm H.c.}\right]. \label{eq:em potential}
\eeq

In order to perform a multipolar expansion, it will be convenient to switch to a description in terms of center-of-mass and relative coordinates, that we shall denote by $\vec R$, $\vec r$ respectively; the  conjugate momenta will be denoted by $\vec P$ and $\vec p$, and we will use $\mu$ for the reduced mass.

\subsection{Power-Zienau-Woolley transformation} \label{subsec:transformation}

In order to study the dynamics induced on the ion by weak fields in the optical regime, we shall use the Dirac-Heisenberg line gauge \cite{Power_Z_1959, BOOK_Schleich_2001}. We shall then perform a multipolar expansion in powers of the small parameter $kr$, where $r$ is the size of the atomic bound state and $k=2\pi/\lambda$ with $\lambda$ the wavelength of the modes under consideration, of the order of the laser wavelength (for the experiment in \cite{Schneider_E_H_S_2010}, $kr \sim 10^{-3}$). 

Starting from Eq.~(\ref{eq:first Hamiltonian}), one obtains the Hamiltonian in the new representation by means of the Power-Zienau-Woolley unitary transformation \cite{Power_Z_1959, P&A_Cohen-Tannoudji_D-R_G_1997}. Although the two representations are equivalent, the minimal coupling Hamiltonian (\ref{eq:first Hamiltonian}) is less suitable to express the problem of the atom interacting with an external field in a truncated basis of eigenstates of the atomic Hamiltonian \cite{Lamb_S_S_1987, Power_T_1978}. In particular, in the Dirac-Heisenberg line gauge the momentum $\vec p$ corresponds, to zero order in the expansion in $kr$, to the mass times the velocity for the relative motion.

We note that this change of representation is commonly used in problems which restrict to the dipolar approximation, and/or the case of neutral particles, and/or assume that the center of mass is fixed~\cite{P&A_Cohen-Tannoudji_D-R_G_1997, BOOK_Schleich_2001, Savolainen_S_1972, Power_Z_1959}. In this paper we deal with a charged particle taking into account the motion of the center of mass and expanding in powers of $kr$ beyond the dipolar approximation, therefore considering a more general framework than the one usually encountered in the literature.

The Power-Zienau-Woolley transformation is defined by the unitary operator $T$, which reads:
\beq
T = {\rm e}^{- {\rm i}S/\hbar},\label{eq:unitary transf:T}
\eeq
where
\beq
S = \int dV \vec P_0 (\vec x) \cdot \vec A_< (\vec x)\,, \label{eq:unitary transf}
\eeq
and the integration is over all space. The polarization operator $\vec P_0$ depends on the atomic variables, and takes the form~\cite{P&A_Cohen-Tannoudji_D-R_G_1997}:
\beq
\vec P_0(\vec x) = \sum_{j=n,e} q_j \big(\vec r_j - \vec R\big) \int_0^1 d\lambda~ \delta\left[\vec x - \vec R -\lambda\big(\vec r_j - \vec R\big)\right], \label{eq:polarization}
\eeq
while $\vec A_<$ represents a vector potential obtained from $\vec A$ in Eq.~(\ref{eq:em potential})
by introducing a high-frequency cutoff in the sum over the modes. This cutoff has to be chosen sufficiently high so that the coupling between the particle and the modes above the cutoff is far off-resonant and thus irrelevant for the processes under study (apart from mass renormalization effects~\cite{Bethe_1947}, which are supposed to be already taken into account).

We now briefly discuss how the operators transform under the unitary transformation defined in Eqs.~(\ref{eq:unitary transf:T})-(\ref{eq:polarization}). The position operators are invariant under the unitary transformation defined by $T$, since they commute with $S$. The momenta, on the other hand, are transformed according to:
\beq
\vec p_j \to T \vec p_j T^\dagger = \vec p_j + \vec\nabla_j S
\eeq
where the gradient is taken with respect to the coordinates of particle $j$. This shows that the direct coupling between the momenta and the vector potential cannot be totally eliminated by this transformation. Nevertheless, to zero order in $kr$ it is possible to suppress the coupling between the relative momentum $\vec p$ and the field. In contrast with the case of a neutral particle, however, in the transformed Hamiltonian the center-of-mass momentum remains coupled to the vector potential $\vec A(\vec R)$.

The form in which the transformation acts on the operators associated to the different fields can be obtained from the commutation relations: $\vec A(\vec x)$ commutes with $\vec A(\xprime)$ and with $\vec B(\xprime)$, so that these operators are invariant under the transformation, while 
\beq
[A_j(\vec x), E_k(\xprime)] = - \frac{i\hbar}{\epsilon_0} \delta_{jk}^\perp (\vec x - \xprime)
\eeq
where $\delta_{jk}^\perp$ is the transverse delta function \cite{P&A_Cohen-Tannoudji_D-R_G_1997}.
This implies that the electric field operator is transformed as follows:
\beq
\vec E \to \vec E' = T \vec E T^\dagger =\vec E - \frac{1}{\epsilon_0} \vec P^\perp_{0<}. \label{eq:E transformed}
\eeq
Here $\vec E'$ is the electric field in the new representation, and $\vec P^\perp_{0<}$ is the polarization of the system of charges as given by (\ref{eq:polarization}), now projected onto its transverse part and including only Fourier modes with wavelengths below the cutoff. The operator $\vec E$ in Eq.~(\ref{eq:E transformed}) can be identified with the electric displacement field (divided by $\epsilon_0$) \cite{P&A_Cohen-Tannoudji_D-R_G_1997}. We shall keep using the notation $\vec E$ to refer to this field, and $\vec E'$ for the electric field.

\subsection{Hamiltonian in Dirac-Heisenberg line gauge} \label{subsec:new Hamiltonian}

After applying the Power-Zienau-Woolley transformation, the Hamiltonian $H'= T H T^\dagger$ reads
\beq
H'= H_{\rm ion} + H_{\rm rad} + H_{\rm coupling} + H_{\rm rel}'.
\eeq
Here $H_{\rm rad}$ has the same form as in Eq.~(\ref{H:rad}) and $H_{\rm rel}'$ contains the relativistic corrections in the new representation, while
\beq
H_{\rm ion} = \frac{\vec P^2}{2M} + \frac{{~\vec p~}^2}{2\mu} + V_{\rm Coul} + E_{\rm dip}
\eeq
depends on the ion degrees of freedom, with $E_{\rm dip}$ the dipolar self-energy:
\beq
E_{\rm dip} = \frac{1}{2\epsilon_0} \int dV (\vec P^\perp_{0<})^2\,,
\eeq
which is a function of the relative coordinate $r$. The remaining Hamiltonian term couples atomic and field degrees of freedom:
\begin{multline}
H_{\rm coupling} = \sum_{j=n,e} q_j \Phi(\vec r_j) + H_{\rm trap} + \frac{\vec p \cdot \vec K_{\rm rel}}{\mu} + \\
+ \frac{\vec P \cdot \vec K_{\rm CM}}{M} + \frac{\vec K_{\rm rel}^2}{2\mu} + \frac{\vec K_{\rm CM}^2}{2M}\,,\label{H:coup}
\end{multline}
where $H_{\rm trap}$ reads
\beq
H_{\rm trap} = -\int dV \vec E_< \cdot \vec P^\perp_{0<}, \label{eq:H-trap}
\eeq
while the other terms in Eq.~(\ref{H:coup}) involve operators defined by:
\begin{align*}
\vec K_{\rm CM} &= \vec K_e + \vec K_n,\\
\vec K_{\rm rel} &= \frac{m_n \vec K_e - m_e \vec K_n}{M},
\end{align*}
with
\beq
\vec K_j = \vec\nabla_j S - q_j \vec A(\vec r_j)\,, \quad j=n,e. \label{eq:Kdef}
\eeq

We shall treat these operators in a multipolar expansion, namely, in powers of $kr$ around the position $\vec R$. The term $H_{\rm trap}$ defined in (\ref{eq:H-trap}) is the one at the basis of the dipole force. To lowest order in $kr$, it corresponds to the dipolar Hamiltonian (\ref{eq:dipolar}), $H_{\rm trap}=-\vec d \cdot \vec E + O(kr)$, where the dipole operator takes now the form
\beq
\vec d = -q_{\rm eff}\vec r \label{eq:dipole}
\eeq
with $q_{\rm eff}$ an effective charge for the dipolar coupling,
\beq
q_{\rm eff} = |q_e| + \frac{m_e}{M} Q. \label{eq:eff charge}
\eeq
The net charge thus introduces a very small correction: For an atomic mass of the order of 10 proton masses, the difference between $q_{\rm eff}$ and $|q_e|$ is of order $10^{-4}$. The other operators read in the multipole expansion:
\bea
\vec K_{\rm CM}^{(0)} &=& -Q \vec A, \label{eq:KCM0}\\
\vec K_{\rm CM}^{(1)} &=& - q_{\rm eff} ~\vec r \times \vec B, \label{eq:KCM1} 
%\vec K_{\rm CM}^{(2)} &\simeq& \frac{q_e}{2} \left(\vec r \cdot \vec \nabla_{\vec R} \right) \left(\vec r \times \vec B\right),
\eea
and
\bea
\vec K_{\rm rel}^{(0)} &=& 0, \label{eq:Krel0}\\
\vec K_{\rm rel}^{(1)} &\simeq& q_e \frac{m_n-m_e}{2M} ~\vec r \times \vec B, \label{eq:Krel1}\\
\vec K_{\rm rel}^{(2)} &\simeq&\frac{q_e}{3} \big(\vec r \cdot \vec \nabla_{\vec R}\big) \big(\vec r \times \vec B\big), \label{eq:Krel2}
\eea
where the superindices indicate the order in $kr$. In Eqs.~(\ref{eq:KCM0})-(\ref{eq:Krel2}), all fields and field derivatives are assumed to be evaluated at the center-of-mass position $\vec R$, and the notation $\vec \nabla_{\vec R}$ stands for a gradient with respect to $\vec R$. We have explicitly reported the terms up to (i) first order in $kr$ and in $m_e/M$, and (ii) second order in $kr$ and zero order in $m_e/M$, as these two small parameters are comparable.

\subsection{Relativistic corrections} \label{subsec:rel corrections}

In the following we shall take into account the relativistic corrections to the Hamiltonian (\ref{eq:first Hamiltonian}), and analyze how the Power-Zienau-Woolley transformation acts on them. In the absence of external fields, the fine structure terms are given by:
\beq
H_{\rm fs} = \frac{1}{2 m_e^2 c^2} \left[-\frac{{\vec p_e~}^4}{4 m_e} + \frac{V_{\rm Coul}'}{r} \vec l \cdot \vec S_e + \frac{\hbar^2}{4} \Delta V_{\rm Coul} \right].
\eeq
We are interested in finding the corrections to these terms that occur when external fields are included. In order to do so, we start from the Dirac equation with a fixed core and apply a Foldy-Wouthuysen transformation, so that the Hamiltonian for the electron including the coupling with the laser field reads in the non-relativistic limit \cite{BOOK_Bjorken_D_1964, Wang_1998}:
\bea
H_{\rm elect} &=& \frac{(\vec p_e-q_e\vec A)^2}{2m_e} + V_{\rm Coul} + q_e \Phi - \frac{q_e\hbar}{2m_e} \vec\sigma \cdot \vec B - \nonumber\\
&& - \frac{(\vec p_e-q_e\vec A)^4}{8m_e^3c^2} - \frac{q_e\hbar}{8m_e^2c^2} \Big\{ i[\vec p_e - q_e\vec A, \vec E_T] + \nonumber\\
&& + \vec\sigma \cdot [\vec E_T \times (\vec p_e-q_e\vec A) - (\vec p_e-q_e\vec A) \times \vec E_T] \Big\} + \nonumber\\
&& + \frac{q_e^2\hbar^2}{8m_e^3c^4} (\vec E_T^2 - c^2 \vec B^2) + \nonumber\\
&& + \frac{e\hbar}{8m_e^3c^2} \{ (\vec p_e-q_e\vec A)^2 , \vec\sigma \cdot \vec B \} + \cdots \label{eq:Dirac_nonrel}
\eea
where $\vec E_T$ denotes the total electric field, including the Coulomb interaction with the core, the external electrostatic field, and the electric field of the laser. In the previous expression, all fields are evaluated at the electron position, and $\{\,,\}$ denotes an anticommutator. The expansion in Eq.~(\ref{eq:Dirac_nonrel}) is in powers of $m_e^{-1}$, which can be interpreted as an expansion in terms of the different momenta compared with $m_ec$. The first three terms in (\ref{eq:Dirac_nonrel}) were already contained in the non-relativistic Hamiltonian (\ref{eq:first Hamiltonian}), while the others give the lowest relativistic corrections.

The Power-Zienau-Woolley unitary transformation acts on the operators in the expression above as explained in subsection \ref{subsec:transformation}; in particular,
\beq
\vec p_e - q_e\vec A ~\to~ \vec p_e +\vec K_e \simeq \vec p + \frac{m_e}{M} \vec P - Q \frac{m_e}{M} \vec A + \frac{1}{2} \vec d \times \vec B \quad
\eeq
where a multipolar expansion has again been performed for the fields, which are then evaluated at the position of the center of mass. The electric field must also be transformed according to eq. (\ref{eq:E transformed}). In this way the extra couplings between the ion and the electromagnetic field that are contained in $H_{\rm rel}'$ can be found. The effect of these terms shall be discussed in Subsection \ref{subsec:spin and fine-structure}.

\section{Efficiency of dipolar trapping of charged particles}
\label{Sec:corrections}

We shall now apply the formalism introduced above in order to determine the efficiency of trapping an alkali-earth metal ion by means of lasers.

We first remark that the dipolar interaction in Eq.~(\ref{eq:H-trap}), in the electric dipole approximation (i.e., in zero order in the expansion in $kr$), is responsible for the optical trapping of the ion, according to the physical processes summarized in Sec.~\ref{Sec:dipolar trapping}. As already mentioned, the presence of a total charge produces a small correction to the effective dipole according to formula (\ref{eq:eff charge}), and hence modifies the potential depth and motional frequencies, as the Rabi frequency $\Omega$ in Eq.~(\ref{eq:Rabi}) now scales with $q_{\rm eff}$. The correction is of the order of the ratio $m_e/M$ between the electronic mass and the total mass of the ion.

With respect to the additional electrostatic field $\vec E_0$, the corresponding coupling with the ion is dominated by the charge monopole, since the typical length scale of this potential is several orders of magnitude larger than the atomic size, and the field intensities involved are too weak to induce significant polarization effects.

The following subsections will be devoted to estimate the relative contributions of the different corrections to the basic dipolar trapping, and point out their similarities and differences in comparison to effects well studied for radiofrequency traps. In order to do so, we shall make reference to the values of the parameters in the experiment reported in \cite{Schneider_E_H_S_2010}. There, a $^{24}$Mg$^+$ ion was optically trapped by a laser with wavelength $\lambda =$ 280 nm. The laser was a tighly focussed Gaussian beam with a waist radius of 7 $\mu$m, which was circularly polarized and red-detuned up to $|\delta| = 2\pi~\times$ 300 GHz with respect to the S$_{1/2}$ $\leftrightarrow$ P$_{3/2}$ transition, with a natural linewidth $\Gamma \sim 2\pi~\times$ 40 MHz. The optical depth, given by the absolute value $|V_{\rm eff}|$ of the effective potential at the minimum of the trap, was $U_0 \lesssim k_B~\times$ 50 mK. The trapping frequencies at the bottom of the trap were $\omega_{\rm radial} \simeq 2\pi~\times$ 200 kHz and $\omega_{\rm axial} \simeq 2\pi~\times$ 2 kHz for the purely optical setup. An electrostatic quadrupole field, with axis at 45$^\circ$ with respect to the trapping laser, provided additional trapping with a frequency of $2\pi~\times$ 45 kHz. Typical frequencies for the oscillatory motion of the atomic center of mass were then of the order of $\omega_0 \sim 2\pi~\times$ 100 kHz, the precise value depending on the direction considered. From these parameters one can identify a hierarchy of frequency scales, displayed in Table \ref{table:frequencies}. The results for the orders of magnitude of the several effects analyzed in the following subsections are summarized in Table \ref{table:corrections}.

\begin{table}[ht]
\begin{tabular}{|c|c|}
\hline
Frequency            	       	& ~Magnitude ($2\pi~\times$ Hz)~ \\ \hline
~$\omega_0,~E_{\rm rec}/\hbar$~	&  (0.5 - 2) $\times~10^5$     	\\ \hline
$\Gamma$                  	& 0.4 $\times~10^8$            	\\ \hline
$U_0/\hbar$               	& 10$^9$           		\\ \hline
$\Omega$                  	& up to 0.3 $\times~10^{11}$   	\\ \hline
$|\delta|$                	& 0.3 $\times~10^{12}$         	\\ \hline
$\omega_L,~\omega_{eg}$        	& $10^{15}$         		\\ \hline
\end{tabular}
\caption{Orders of magnitude for the different frequency scales involved in the problem, taking parameters from \cite{Schneider_E_H_S_2010}. The symbols have been defined in the text. \label{table:frequencies}}
\end{table}

\begin{table}[ht]
\begin{tabular}{|c|c|c|}
\hline
Effect            	       			& ~$\delta E/U_0$~& ~Subsection~			\\ \hline
$Q$-dependent correction to $\vec d$ 		& $10^{-4}$	& \ref{subsec:new Hamiltonian}		\\ \hline
~Coupling in relativistic corrections~		& $10^{-5}$ 	& \ref{subsec:spin and fine-structure}	\\ \hline
Higher-order multipole terms			& $10^{-6}$ 	& \ref{subsec:multipole}		\\ \hline
Monopole coupling $-Q \vec P\cdot\vec A/M$ 	& $10^{-8}$	& \ref{subsec:total-momentum} 		\\ \hline
\end{tabular}
\caption{Orders of magnitude estimated for the most important corrections to the dipolar trapping Hamiltonian. We refer to the subsections indicated for further details. Much smaller effects have been found as a result of the monopole coupling with blackbody radiation (which heats the motion at a rate $\Gamma'\approx10^{-7}$ Hz, as shown in Subsection \ref{subsec:thermal rad}), and of the time-dependence of the optical potential (resulting in a micromotion with amplitude $10^{-20}$ times smaller than the secular motion of the center of mass, as seen in Subsection \ref{subsec:micromotion}). \label{table:corrections}}
\end{table}

\subsection{Coupling of the electric field to the total charge} \label{subsec:total-momentum}

Our primary interest is the study of effects due to the net charge of the trapped particle. Thus, we shall consider first the term in $H_{\rm coupling}$ in which the monopole appears to the lowest order in the multipolar expansion, namely, the term that couples the field with the center-of-mass momentum $\vec P$. To first order in $kr$, it takes the form:
\beq
\frac{\vec P}{M} \cdot \vec K_{\rm CM} \simeq \frac{\vec P}{M} \cdot \left( - Q \vec A + \vec d \times \vec B \right)\,. \label{eq:P coupling}
\eeq
The first term is in zero order in $kr$ and describes the coupling between a charge monopole and the electromagnetic field, while the second term is in first order and involves both the center-of-mass and the dipolar degrees of freedom. 

In order to estimate the magnitude of the zero-order term we take the following approach: We focus on the center-of-mass motion and make a harmonic approximation for the effective potential $V_{\rm eff}(\vec R)$, namely, we make a second-order Taylor expansion about the equilibrium position of the center of mass. This system has then an effective Hamiltonian:
\beq
H_{\rm eff} = \frac{(\vec P-Q\vec A)^2}{2M} + \frac{M\omega_0^2\vec R^2}{2} \label{eq:H_eff_COM_motion}
\eeq
where $\vec A$ is the vector potential for the laser field, and where the motional frequency $\omega_0$ is given, for the purely optical trapping, by the laser intensity and the waist of the beam at the focus. For the sake of simplicity we have assumed the same trapping frequency for all spatial directions, since we only wish to make an order-of-magnitude estimation. To this effect, we shall take $\omega_0$ of about ten orders of magnitude smaller than the frequency of the laser field. In the previous effective Hamiltonian (\ref{eq:H_eff_COM_motion}) we have also included the term $(Q\vec A)^2/(2M)$, the zero-order contribution to the terms in $H_{\rm coupling}$ which are quadratic in the field operators. Equation (\ref{eq:H_eff_COM_motion}) describes a harmonic oscillator driven by a field with frequency $\omega_L\gg\omega_0$. The effect of the coupling is then to induce a micromotion-like oscillation with very small amplitude at the frequency of the laser. Typical energies associated to this oscillation are of the order of $(Q \vec A)^2/(2M)$, which is about eight orders of magnitude smaller than the optical depth, and four orders smaller than the motional energy scale given by $\hbar \omega_0$ (we note that in conventional rf-traps the energies associated to secular motion and micromotion are normally of the same order of magnitude). 

The following term in the coupling to the total momentum (\ref{eq:P coupling}) is of order $P/(Mc)\lesssim10^{-8}$ when compared to the dipolar trapping term (\ref{eq:dipolar}). Besides, since $\vec B$ and $\vec E$ are out of phase, this term is proportional to the component of the dipole oscillation that is out of phase with respect to the electric field, and which is smaller by a factor $\Gamma/\delta \sim 10^{-4}$ with respect to the in-phase component of the dipole oscillation, associated with the optical trapping. Taking both aspects into account, this term can then be neglected.

\subsection{Time-dependence of the effective potential} \label{subsec:micromotion}

The effective potential $V_{\rm eff} (\vec R)$ for the center of mass results from averaging the optical potential over the fast oscillation period of the laser field. Indeed, the time-dependent optical potential experienced by the center of mass can be written as:
\begin{align}
V_{\rm optical}(\vec R, t) &=  2 V_{\rm eff}(\vec R) \cos^2(\omega_Lt) = \nonumber\\
& = V_{\rm eff}(\vec R) [1 + \cos(2\omega_Lt)], \label{eq:time-dep potential}
\end{align}
as the dipolar force is proportional to the product of the atomic dipole and the electric field, each of them oscillating at the optical frequency $\omega_L$. One might wonder whether by including the time dependence of this potential some other relevant micromotion-like effect could appear. 

Once again we can estimate the importance of this effect by considering a harmonic approximation with trapping frequency $\omega_0$ for $V_{\rm eff}(\vec R)$. Under this approximation, the time-dependent optical potential (\ref{eq:time-dep potential}) is of the same form as the one for the motion of an ion in each of the transverse directions in a Paul trap:
\beq
V_{\rm Paul}(x) = \frac{Mx^2}{2} \frac{\omega_{RF}^2}{4} [a - 2q \cos(\omega_{RF} t)],
\eeq
which leads to a Mathieu equation for the coordinate $x$. By comparing the two previous potentials one finds that the parameters $a$ and $q$ for the dipole trap are both of the order of $(\omega_0/\omega_L)^2 \sim 10^{-20}$. This quantity determines the order of magnitude for the ratio of the high-frequency components with respect to the secular motion in the solution of the Mathieu equation (and the parameters are in the stability region since, for the purely optical potential, $a>0$) \cite{BOOK_Abramowitz_S}. The micromotion amplitude associated with the time dependence of the optical potential is thus negligible. We note that because in this case both parameters $a$ and $q$ scale as $(\omega_0/\omega_L)^2$, the kinetic contribution coming from the micromotion can also be safely ignored (as opposed to the results discussed in \cite{Cirac_G_B_P_Z_1994} for a trapped ion with $a=0$ and $q\to0$ as $\omega_{RF}\to\infty$, since in the case $a=0$ the micromotion's kinetic energy has a different scaling). 

The previous considerations have taken only the optical potential into account. In a setup as the one in Fig. \ref{fig:trap}, in which a static quadrupole field $\vec E_0$ is also included, this field will modify the value of the parameter $a$ in the corresponding Mathieu equation (because the static trapping potential along the propagation direction of the laser is obtained at the cost of a repulsive transverse force). However, as long as this electrostatic force is weak enough compared to the dipole force, as was the case in the experiment reported in \cite{Schneider_E_H_S_2010}, the conclusions in the previous paragraph remain valid. Finally, it is worth noting that the micromotion frequency resulting from the time dependence of the potential (\ref{eq:time-dep potential}) is of about $2\omega_L$, and thus far-off resonance from the force exerted by the laser on the net charge (which was analyzed in subsection \ref{subsec:total-momentum} and has frequency $\omega_L$).

\subsection{Coupling between the center-of-mass motion and the blackbody radiation} \label{subsec:thermal rad}

The coupling between the net charge and the external field could become important if field modes with frequencies similar to the motional frequencies of the center of mass are taken into account. To study this effect, which could give rise to additional heating mechanisms, we shall consider once more the motion of the center of mass in harmonic approximation for the effective potential, and analyze the coupling between this charged harmonic oscillator and the continuum of modes of the electromagnetic field.
This problem can be modelled by the Hamiltonian $H_{\rm eff} + H_{\rm rad}$ with $H_{\rm eff}$ introduced in Eq. (\ref{eq:H_eff_COM_motion}), and $H_{\rm rad}$ defined in (\ref{H:rad}).
Following \cite{Schneider_E_H_S_2010}, we shall be interested in motional trapping frequencies ranging from 10 kHz and 1MHz. For these values, the wavelength of the field modes which are close to resonance is many orders of magnitude larger than the typical size of the atomic motion, and then the field can be treated as spatially constant.

We wish to estimate the rates at which energy is exchanged between the harmonic oscillator and the continuum of modes, assuming that the electromagnetic field is in a thermal state. This calculation is carried out (for instance) in \cite{API_Cohen-Tannoudji_D-R_G_1998}, where the rate of heating of the motion due to the interaction with the thermal bath is shown to be given by a quantity $\Gamma'$ associated to stimulated transitions, and thus dependent on $\langle n(\omega_0)\rangle$, the mean population of the modes of the continuum which are resonant with the motion (the rate of energy exchange also depends on the energy of the charged particle, but this is negligigle in our problem). For $\omega_0$ of the order of $2\pi~\times$ 100 kHz, the quantity $\hbar\omega_0/k_B$ corresponds to a temperature of a few $\mu$K. The mean population of the modes resonant with $\omega_0$ for a radiation bath at room temperature $T$ is thus $\langle n (\omega_0)\rangle \approx (k_BT)/(\hbar\omega_0) \approx 10^8$. The resulting heating rate is $\Gamma'\approx10^{-7}$ Hz, and the associated time scale is of a few months. \footnote{One can also check whether, even if the heating rate is very low, the energy shift of the levels of the harmonic oscillator due to the coupling with the electromagnetic field can be significant. Along the lines of \cite{Bethe_1947}, in order to avoid divergences a renormalization term can be included accounting for the energy shift of a free particle, and then a relativistic cutoff of order $Mc^2$ is introduced in the coupling. The result for the energy shift is of about twenty orders of magnitude smaller than the zero-point energy $\hbar\omega_0$ of the harmonic approximation for the motion.}

\subsection{Higher-order corrections in the coupling between the laser and the relative motion} \label{subsec:multipole}

As seen in the previous subsections, the monopole coupling between the center of mass and the electromagnetic field gives rise to very small effects. We shall then study the dominant higher-order corrections to the dipolar interaction which couples the electronic degrees of freedom and the field, and conclude that the most important corrections are also present when considering the dipolar coupling of a neutral atom, so that no relevant additional effects that depend on the total charge of the particle appear.

\subsubsection{Multipole expansion of $H_{\rm trap}$} 

We first focus on Hamiltonian (\ref{eq:H-trap}), which at lowest order in the expansion in $kr$ (electric dipole approximation) determines the optical potential, and discuss now the higher-order corrections in powers of $kr$, taking the lowest order in the expansion in powers of the ratio $m_e/M$. The electric quadrupole correction reads
\beq
H_{\rm trap}^{(1)} \simeq -\frac{q_e}{2} ~ \big(\vec r \cdot \grad_{\vec R}\big) \big(\vec r \cdot \vec E\big). \label{eq:Htrap 1st correction}
\eeq
This term is far-off resonance: The operator $r^2$ does not connect the levels involved in the dipolar transition driven by $E$, because they have different parity; the non-vanishing matrix elements in $H_{\rm trap}^{(1)}$ will thus oscillate too fast to contribute significantly to the evolution. Indeed, the effect of the coupling (\ref{eq:Htrap 1st correction}) can be estimated by means of time-dependent perturbation theory. First-order processes will give rise to transitions in the dressed-state basis with probability amplitudes of the order of $(kr)\Omega/\omega_L\sim 10^{-8}$, and second order processes will have associated rates that scale like $(kr)^2\Omega^2/\omega_L$. These contributions are then smaller than the following order in the expansion of $H_{\rm trap}$ in powers of $kr$, which we denote by $H_{\rm trap}^{(2)}$. The dominant correction to the dipolar coupling is then given by the electric octupole of (\ref{eq:H-trap}),
\beq
H_{\rm trap}^{(2)} \simeq -\frac{q_e}{6} ~ \big(\vec r \cdot \grad_{\vec R}\big)^2 \big(\vec r \cdot \vec E\big), \label{eq:H-trap correction}
\eeq
which is of order of $(kr)^2$ compared to the zero-order term $-\vec d\cdot\vec E$.

\subsubsection{Coupling of the field with the relative momentum} 

We next analyze the term in $H_{\rm coupling}$, Eq.~(\ref{H:coup}), which describes the coupling of the external field with the relative momentum, $\vec p\cdot \vec K_{\rm rel}/\mu$, with $\vec K_{\rm rel}$ reported in Eqs.~(\ref{eq:Krel0})-(\ref{eq:Krel2}). In lowest order in $kr$, and neglecting terms of order $(kr)(m_e/M)^2$, it reads
\beq
\frac{\vec p \cdot \vec K_{\rm rel}^{(1)}}{\mu}
\simeq
-\frac{q_e}{2} \left(\frac{1}{m_e}-\frac{1}{m_n}\right) \vec l \cdot \vec B\,, \label{eq:r-B}
\eeq
where $\vec l=\vec r\times \vec p$ is the orbital angular momentum of the electron. This term describes the coupling between the magnetic dipole, induced by the orbital motion of the valence electron, and the magnetic field, and it connects bound states with equal parity. It is thus far-off resonance from the transitions it couples, and hence negligible. In presence of static magnetic fields, it may give rise to energy shifts whose magnitude depends on the angular momentum.

The following terms in powers of $kr$ have resonant contributions which must be estimated. The dominant part (i.e., to leading order in $m_e/M$) is given by the expression:
\beq
\vec p \cdot \frac{\vec K_{\rm rel}^{(2)}}{2\mu} \simeq -\frac{q_e}{3 m_e} \big(\vec r \cdot \grad_{\vec R}\big) \big(\vec l \cdot \vec B\big).
\eeq
Similarly to (\ref{eq:H-trap correction}), this term is of order $(kr)^2$ when compared to the dipolar coupling, and to this order it does not depend on the total charge $Q$.

\subsubsection{Effect of the previous corrections}

The terms so far discussed constitute corrections to the coupling between electric field and atomic dipole, so that the Rabi frequency defined in (\ref{eq:Rabi}) is modified according to:
\begin{multline}
\Omega(\vec R) = \left|q_{\rm eff} \bra{g} \vec r \cdot \vec E \ket{e}\right| \to
\Bigg| q_{\rm eff} \bra{g} \vec r \cdot \vec E \ket{e} - \\
- \frac{q_e}{6} \bra{g} (\vec r \cdot \grad_{\vec R}) \! \left[ (\vec r \cdot \grad_{\vec R}) (\vec r \cdot \vec E) + \frac{2}{m_e} (\vec l \cdot \vec B) \right] \! \ket{e}\Bigg|,
\end{multline}
which is reported in rotating-wave approximation and up to second order in $kr$ and first order in $(kr)(m_e/M)$. As a consequence, the resulting dipolar potential, that depends on the Rabi frequency as shown in formulas (\ref{eq:s})-(\ref{eq:effective potential}) can be written as $V_{\rm eff}=V_{\rm eff}^{(0)}+V_{\rm eff}^{(2)}$, where the superscript indicates the order in $kr$, with
\beq
V_{\rm eff}^{(0)} \propto |q_{\rm eff} \bra{g} \vec r \cdot \vec E \ket{e}|^2,
\eeq
the dipolar potential in the electric dipole approximation, and
\begin{multline}
V_{\rm eff}^{(2)} \propto \frac{q_e^2}{3} {\rm Re}\Big\{\bra{g} \vec r \cdot \vec E \ket{e}^* \bra{g} \Big[ (\vec r \cdot \grad_{\vec R})^2 (\vec r \cdot \vec E) + \\
+ \frac{2}{m_e} (\vec r \cdot \grad_{\vec R}) (\vec l \cdot \vec B) \Big] \ket{e} \Big\} \label{eq:potential_correction}
\end{multline}
the first correction in the multipole expansion. Its order of magnitude can be estimated by considering that it can be rewritten in terms of second derivatives of the electric field, and hence in terms of the typical length scale over which the electric field changes.  In particular, if the electric field is sufficiently smooth in the transverse plane, then the dominant corrections are due to the derivatives along the axial direction, which are proportional to $k_L$. The resulting terms in $V_{\rm eff}^{(2)}$ will then be proportional to the light intensity, just as $V_{\rm eff}^{(0)}$, and will only provide a global factor giving a relative difference of order $(kr)^2 \sim 10^{-6}$ in the values of the potential depth $U_0$ and the motional frequencies with respect to the ones obtained from the lowest-order term in the potential. If instead the transverse derivatives are not negligible, then the corrections arising from them may slightly modify the shape of the effective potential (an effect which can be at most of the same order of magnitude as the one related to the axial derivatives).

\subsubsection{Terms quadratic in the field operators}

Finally, there are some additional terms in $H_{\rm coupling}$ which are quadratic in the field. The zero-order in $kr$, of the form $(Q\vec A)^2/(2M)$, couples only to the center of mass and has been discussed in Subsection \ref{subsec:total-momentum}. The next two terms in the expansion in powers of $kr$ are (to lowest order in $m_e/M$) given by:
\beq
-\frac{Q}{M}\big(\vec d\times\vec B\big)\cdot \vec A + \frac{1}{8\mu} \big(\vec d\times\vec B\big)^2.
\eeq
These two couplings are in principle of the same order, but the first of them is off-resonant from the transitions it couples, so that the dominant contribution will be given by the second one. In a two-level description of the dipolar trapping, the latter term corresponds to a diagonal operator which induces different energy shifts in the states $\ket{g}$, $\ket{e}$, thus modifying the frequency of the atomic transition. For the magnetic field of the trapping laser, the order of magnitude of this intensity-dependent shift is of about 1 Hz, which means nine orders of magnitude smaller than the shift associated to the potential depth. This effect is thus smaller than the ones found previously, and it is also independent of the total charge.

\subsection{Spin coupling and corrections to the fine structure terms} \label{subsec:spin and fine-structure}

To conclude our analysis, we consider the couplings coming from the relativistic terms in $H_{\rm rel}'$, obtained in Subsection \ref{subsec:rel corrections}.
Typical experiments include weak magnetic fields, used to set a preferred quantization axis for the angular momentum. The magnetic field of the laser could in principle induce transitions between different states and thus interfere with the dipole trap-mechanism. However, the spin coupling, given by the term
\beq
H_{\rm spin} = - \frac{q_e\hbar}{2m_e} \vec\sigma \cdot \vec B
\eeq
in the Hamiltonian $H_{\rm rel}'$, oscillates with the optical frequency, and then much faster than all the other relevant frequencies involved. The probability for spin transitions in first-order perturbation theory scales as $(g_e \mu_B B_L)^2 / (\hbar \omega_L)^2 \sim 10^{-15}$ (where the subscript $L$ indicates that the magnetic field corresponds to the trapping laser). Second-order processes have associated rates of the order of $\sim 10^{-15}\omega_L$, and thus very slow compared to the other dynamical phenomena under study.

The relevance of the remaining relativistic corrections to the coupling between the laser field and the ion can be estimated by plugging orders of magnitude and using parity and resonance arguments, as has been done in the previous subsections. The largest term comes from the presence of the electric field of the laser in the spin-orbit coupling in the third line of eq. (\ref{eq:Dirac_nonrel}), which gives rise to an interaction of the form:
\beq
- \frac{q_e\hbar}{(2m_ec)^2} \vec\sigma \cdot \big(\vec E_L \times \vec p~\big).
\eeq
This extra coupling can be added to the dipolar Hamiltonian (\ref{eq:dipolar}), causing a spin-dependent modification of the coupling constant of order at most $10^{-5}$. As was the case with the corrections found in the previous subsections, this is a charge-independent effect, namely, it is present also in the trapping of neutral atoms.

\section{Final remarks}
\label{Sec:conclusions}

We have studied the trapping of an ionized atom with one valence electron by means of weak optical fields. The different terms that couple the ion and the field have been analyzed, and the effects due to the net charge of the particle have been identified. From the estimations carried out in this work, one can conclude that the couplings relevant for the dipolar trapping of a charged particle are to very good approximation equivalent to those present for neutral particles. The most important effect due to the charge monopole is associated with the direct coupling between the vector potential of the external field and the center-of-mass momentum of the charged particle. For the field of the trapping laser, this term gives rise to a micromotion-like effect with energy of order 10$^{-8}$ with respect to the potential depth of the dipolar trap, as found in subsection \ref{subsec:total-momentum}. The net charge can also be affected by other external electric fields, such as the one used in \cite{Schneider_E_H_S_2010} to confine the particles in the axial direction. This implies the need to carefully suppress undesired additional fields, specially close to resonance with the center-of-mass motion; however, only dc-components of the stray fields get compensated up to now. Blackbody radiation at room temperature, nevertheless, does not couple strongly enough to the charge so as to produce significant heating in the time scales of the problem. 

One interesting aspect that remains open is connected with the minimum temperatures achievable in dipole traps when ions and atoms are trapped together. Indeed, one of the motivations for the optical trapping of ions is the possibility to use the same setup to confine both neutral and charged particles forming composite systems. Such hybrid systems have already been experimentally realized with ions in Paul traps, allowing for the observation of cold ion-atom collisions \cite{Grier_C_O_V_2009} and sympathetic cooling of an ion by a Bose-Einstein condensate \cite{Zipkes_P_S_K_2010}. However, the micromotion of the ions in the Paul trap represents a limitation to the range of energies that can be explored \cite{Zipkes_R_S_K_2010,Schmid_2010}. The possibility of optical trapping of ions is then an attractive alternative, since the micromotion of an ion in a dipolar trap is much smaller than in rf traps. 

Nevertheless, even in purely optical traps, only the charged particles will be subject to the extra driving due to the coupling of the monopole with the field. The effect of this coupling has been studied in Subsection \ref{subsec:total-momentum}, and the kinetic energies associated to it were estimated to be of about $(Q \vec A)^2/(2M)$. This phenomenon can thus give rise to atom-ion collisions in a similar way as in the case of Paul-trap setups, but with an energy scale which is smaller by several orders of magnitude: Micromotion in radiofrequency traps has associated energy scales of the order of $k_B$ times a few $\mu$K (for very cold ions and good micromotion compensation), whereas the driven motion in the dipolar trap for the setup of \cite{Schneider_E_H_S_2010} can be estimated to have a kinetic energy smaller than $k_B ~\times$ 1 nK. We note that for red-detuned trapping beams both the kinetic energy associated to this micromotion and the trapping depth are proportional to the laser intensity; therefore, if the laser intensity is increased to achieve larger trapping depths the energy associated to the driven oscillation will increase by the same amount. However, for blue-detuned trapping the particles are confined around the nodes of the field, and thus this micromotion-like effect can be further suppressed. These considerations indicate that laser trapping, for instance in optical lattices, is a promising approach for realizing hybrid ultracold atom-ion systems \cite{Doerk_I_C_2010}.

The realization of composite charged-neutral systems can allow for the study of impurities in a bath \cite{Goold_et_al_2010} and ultracold atom-ion collisions \cite{Idziaszek_et_al_2009}. The possibility to create optical potentials for ions has also been proposed for the implementation of Frenkel-Kontorova models \cite{Garcia-Mata_Z_S_2007}. We expect our work to provide a useful reference for an accurate description of the corresponding dynamics.

\acknowledgements

This work has been partially supported by the European Commission (Integrating Project AQUTE, STREP PICC), by the Alexander-von-Humboldt Foundation, and by the German Research Foundation (Heisenberg professorship, MO1845/1-1).

\end{document}